\documentclass[conference]{IEEEtran}
\usepackage{graphicx}
\usepackage{amsmath}

\ifCLASSINFOpdf
\else
\fi
\begin{document}
%
\title{Optimal portfolio for a robust financial system}

\author{
\IEEEauthorblockN{Yoshiharu Maeno\\ and Satoshi Morinaga}
\IEEEauthorblockA{NEC Corporation\\
Kawasaki-shi, Kanagawa 211-8666, Japan\\
Email: y-maeno@aj.jp.nec.com\\
Email: morinaga@cw.jp.nec.com}
\and
\IEEEauthorblockN{Kenji Nishiguchi}
\IEEEauthorblockA{Japan Research Institute\\
Shinagawa-ku, Tokyo 141-0022, Japan\\
Email: nishiguchi.kenji@jri.co.jp}
\and
\IEEEauthorblockN{Hirokazu Matsushima}
\IEEEauthorblockA{Institute for International\\Socio-economic Studies\\
Minato-ku, Tokyo 108-0073, Japan\\
Email: h-matsushima@ah.jp.nec.com}
}

%


\maketitle

\begin{abstract}
This study presents an ANWSER model (asset network systemic risk model) to quantify the risk of financial contagion which manifests itself in a financial crisis. The transmission of financial distress is governed by a heterogeneous interbank credit network and an investment portfolio of banks. Bankruptcy reproductive ratio of a financial system is computed as a function of the diversity and risk exposure of an investment portfolio of banks, and the denseness and concentration of a heterogeneous interbank credit network. An analytic solution of the bankruptcy reproductive ratio for a small financial system is derived and a numerical solution for a large financial system is obtained. For a large financial system, large diversity among banks in the investment portfolio makes financial contagion more damaging on the average. But large diversity is essentially effective in eliminating the risk of financial contagion in the worst case of financial crisis scenarios. A bank-unique specialization portfolio is more suitable than a uniform diversification portfolio and a system-wide specialization portfolio in strengthening the robustness of a financial system.
\end{abstract}


%
\IEEEpeerreviewmaketitle

\section{Introduction}

The recent European sovereign debt crisis has impaired many European banks. The financial distress on some European banks may transmit to banks in other continents and cause a catastrophic financial contagion worldwide. Indeed, more than 10 banks in US, UK, and other countries went bankrupt in the financial crisis ensuing from the collapse of the subprime mortgage market in 2007. Since then, supervisors and other relevant authorities have directed a great effort at comprehending the systemic risk hidden behind banks and finding the solution to eliminate financial contagion.

Recently, Monte-Carlo simulation models were developed to understand the transmission of distress on banks and quantify the risk of financial contagion\cite{Arinaminpathy}, \cite{Gai11}, \cite{Gai}, \cite{Jaramillo}, \cite{Gatti}. Two categories of financial distress were studied with the models. One is a defective investment portfolio of banks which weakens the robustness of a financial system when the market prices fluctuate\cite{Beale}. The other is evaporating solvency of failing banks in paying off interbank borrowings\cite{Haldane}, \cite{Upper}, \cite{May}, \cite{Nier}. Both categories of financial distress transmit simultaneously in the financial crisis although either category may transmit separately in the peace time. None of the models address the compound financial distress of both categories. None of the models are founded solidly on the stylized facts for the statistics of the funds transfer between banks\cite{Becher}, \cite{Soramaki}, \cite{Inaoka} and the process where a bank go bankrupt\cite{Duffie} either. These are the issues of the models.

This study presents an ANWSER model (asset network systemic risk model) to quantify the risk of financial contagion in a financial crisis. The ANWSER model analyzes the transmission of both categories of financial distress. The transmission of financial distress is governed by a heterogeneous interbank credit network and an investment portfolio of banks. The ANWSER model computes a bankruptcy reproductive ratio of a financial system as a function of the diversity and risk exposure of an investment portfolio of banks, and the denseness and concentration of a heterogeneous interbank credit network.

The ANWSER model is described in \ref{ANWSER model}, including the balance sheet of banks, a interbank credit network, an investment portfolio of banks, the details of a financial crisis scenario, and the definition of the bankruptcy reproductive ratio. An analytic solution of the bankruptcy reproductive ratio for a small financial system is derived in \ref{Analytic solution for $N=2$} and a numerical solution for a large financial system is obtained in \ref{Numerical solution for $N=500$}. Risk landscape of the bankruptcy reproductive ratio is drawn on the plane of the diversity and risk exposure of an investment portfolio of banks. The risk landscape can be a basic compass tool in searching for the optimal portfolio which strengthens the robustness of a financial system.

\section{ANWSER model}
\label{ANWSER model}

\subsection{Balance sheet}
\label{Balance sheet}

\begin{figure}[!t]
\centering
\includegraphics[width=1.5in, angle=-90]{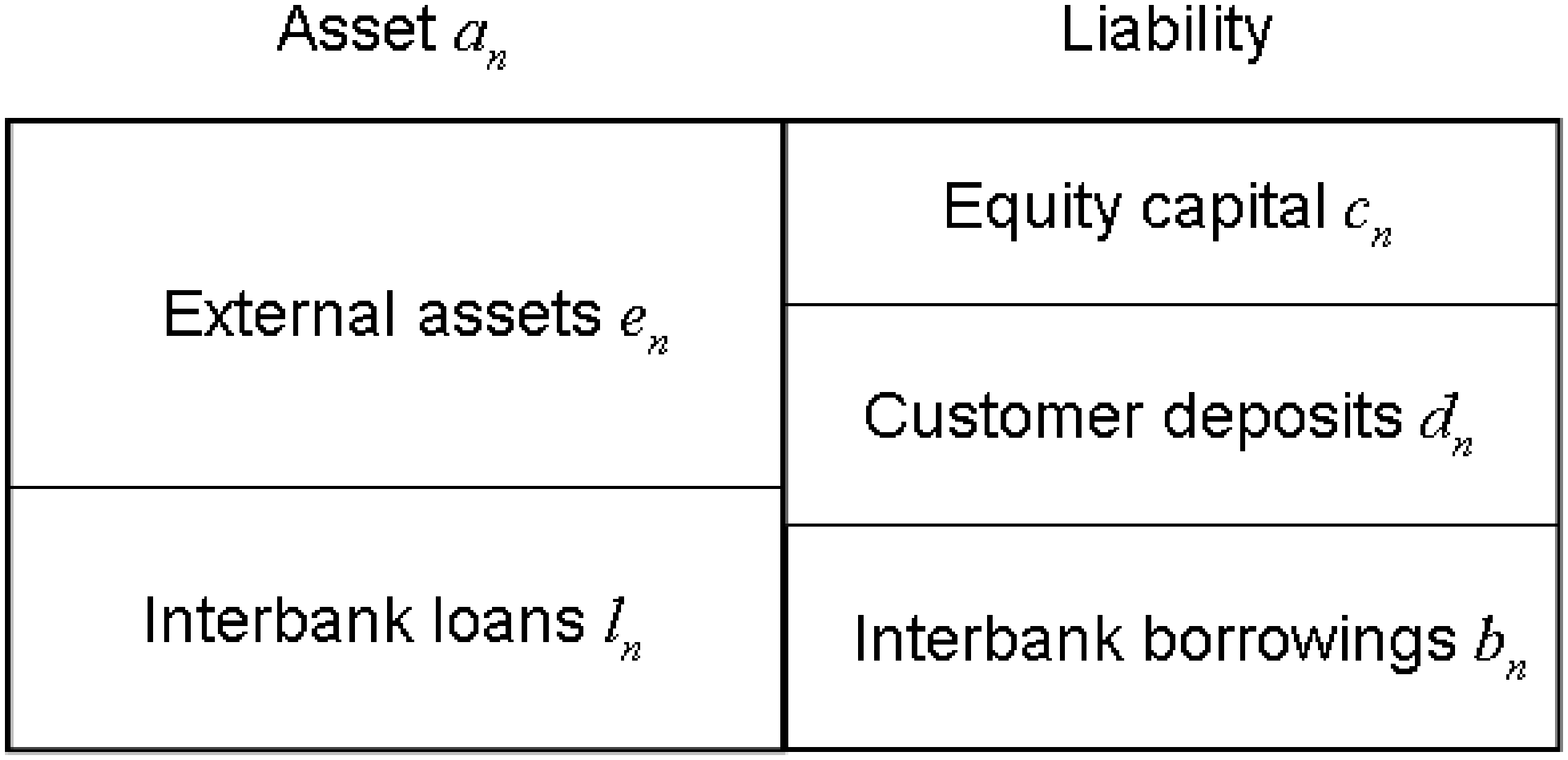}
\caption{Balance sheet of the $n$-th bank in the ANWSER model.}
\label{1124b1}
\end{figure}

The balance sheet of a bank is represented by a set of five quantities in the ANWSER model. The amount of asset of the $n$-th bank is $a_{n}$. The number of banks is $N$ ($1 \leq n \leq N$). The asset include external assets $e_{i}$ and interbank loans $l_{n}$. The external asset is an investment in general. The total amount of interbank loans is given by $L = \sum_{n=1}^{N} l_{n}$. The liability includes equity capital $c_{n}$, interbank borrowings $b_{n}$, and deposits $d_{n}$. The equity capital means the core tier 1 capital including common stock and disclosed reserves. These need not be paid off and can be used to absorb the loss from the distress immediately. An interbank borrowing of one bank is an interbank loan of other bank. Fig.\ref{1124b1} shows the balance sheet.

The interbank loan ratio $\theta$ of the financial system is defined by Eq.(\ref{thetadef}).
\begin{equation}
\theta = \frac{L}{\sum_{n=1}^{N} a_{n}}.
\label{thetadef}
\end{equation}

The equity capital ratio $\gamma$ of individual banks is defined by Eq.(\ref{gammadef}).
\begin{equation}
\gamma_{n} = \frac{c_{n}}{a_{n}}.
\label{gammadef}
\end{equation}

It is assumed in this study that big banks and small banks have the same equity capital ratio. This value, $\gamma$, is the minimal level of the equity capital ratio required by the bank regulatory policies.

\subsection{interbank credit network}
\label{interbank credit network}

An interbank loan is the credit relation between a creditor bank and a debtor bank which appears when the debtor bank raises money in the interbank market. A interbank credit network describes the all credit relations. It is a directed graph which consists of banks as vertices, and the interbank loans as edges from creditor banks to debtor banks. The interbank credit network is specified by an $N \times N$ matrix $\mbox{\boldmath{$T$}}$ in the ANWSER model. If the $n$-th bank lends $n'$-th bank, $T_{nn'}=1$ and 0 otherwise. The outgoing degree of the $n$-th bank is the number of banks which it lends. It is calculated by $k_{n}^{{\rm out}} = \sum_{n'=1}^{N} T_{nn'}$. The incoming degree is the number of banks from which it borrows. It is calculated by $k_{n'}^{{\rm in}} = \sum_{n=1}^{N} T_{nn'}$.

It is found from the studies on the statistics of some electronic funds transfer system like BOJ-Net in Japan\cite{Inaoka}, Fedwire in US\cite{Soramaki}, CHAPS in UK\cite{Becher}, and e-MID in Italy\cite{Masi} that the distribution of the degree of banks obeys a power law. And more importantly, the amount of funds transferred between banks increases as a power of their degrees. It is assumed in the ANWSER model that the generalized law in eq.(\ref{wnn'}) holds. The amount of an interbank loan from the $n$-th bank to the $n'$-th bank is $w_{nn'}$\cite{Maeno}.
\begin{equation}
w_{nn'} = \frac{ T_{nn'} \, (k_{n}^{{\rm out}} \, k^{{\rm in}}_{n'})^r }{\sum_{n \neq n'} T_{nn'} \, (k_{n}^{{\rm out}} \, k^{{\rm in}}_{n'})^r } L.
\label{wnn'}
\end{equation}

In the ANWSER model, the parameter $r \geq 0$ governs the heterogeneity of the distribution of interbank loans over the interbank credit network. The amount of interbank loans is constant if $r=0$. The distribution becomes more heterogeneous as $r$ increases. Given $\mbox{\boldmath{$T$}}$ and $r$, the amount of interbank loans and borrowings of individual banks is determined by eq.(\ref{ln}) and (\ref{bn'}).
\begin{equation}
l_{n} =\sum_{n' \neq n} w_{nn'}.
\label{ln}
\end{equation}
\begin{equation}
b_{n'} = \sum_{n \neq n'} w_{nn'}.
\label{bn'}
\end{equation}

Given $\theta$, the amount of external assets is determined by eq.(\ref{en}). A prerequisite that the external assets are no less than the net interbank borrowings is imposed because the bank has already gone bankruptcy if this prerequisite is not satisfied.
\begin{equation}
e_{n} = \max(b_{n}-l_{n}, 0) + (\frac{1-\theta}{\theta}L - \sum_{n=1}^{N} \max(b_{n}-l_{n}, 0)) \frac{l_{n}}{L}.
\label{en}
\end{equation}

Given $\gamma$, $c_{n}=\gamma(e_{n}+l_{n})$ and $d_{n}=(e_{n}+l_{n})-(c_{n}+b_{n})$. The value of the five quantities in the balance sheet in \ref{Balance sheet} is determined.

Two macroscopic indices are defined to summarize the denseness and concentration of a interbank credit network. The average degree $\kappa$ is defined by eq.(\ref{kdef}). $0 < \kappa \leq N-1$ is satisfied. A more dense interbank credit network has a larger value of $\kappa$. It is a function of $\mbox{\boldmath{$T$}}$.
\begin{equation}
\kappa(\mbox{\boldmath{$T$}}) = \frac{1}{N(N-1)} \sum_{n \neq n'} T_{nn'}.
\label{kdef}
\end{equation}

The share $\rho_{5}$ of the largest five banks in the interbank loans is defined by eq.(\ref{R5def}). Here, $n_{1}$ refers to the bank which owns the largest amount of interbank loans, $n_{1} = \arg \max_{n} (l_{n})$, and $n_{i}$ refers to the bank which owns the $i$-th largest interbank loans. $0 \leq \rho_{5} \leq 1$ holds. A more concentrated interbank credit network has a larger value of $\rho_{5}$. The index $\rho_{5}$ is an increasing function of $r$.
\begin{equation}
\rho_{5}(r) = \frac{\sum_{n=n_{1}, n_{2}, n_{3}, n_{4}, n_{5}} l_{n}}{\sum_{n=1}^{N} l_{n}}.
\label{R5def}
\end{equation}

\subsection{Investment portfolio}
\label{Investment portfolio}

The external assets of a bank are the investment in assets like securities and government bonds. $X_{nm}$ is the amount of the $m$-th asset as a fraction of the external assets which the $n$-th bank invests. The number of assets is $M$ ($1 \leq m \leq M$). $\sum_{m=1}^{M} X_{nm}=1$ for any $n$ and $0 \leq X_{nm} \leq 1$ are satisfied. The investment portfolio is specified by an $N \times M$ matrix $\mbox{\boldmath{$X$}}$ in the ANWSER model.

Two macroscopic indices are defined to summarize the diversity and risk exposure of the investment portfolio\cite{Beale}. The index $\delta$ is the average difference in the fractions between banks, $X_{nm}-X_{n'm}$. It is defined by eq.(\ref{Ddef}). The relation $0 \leq \delta \leq 1$ is satisfied. A more diverse investment portfolio has a larger value of $\delta$. It is a function of $\mbox{\boldmath{$X$}}$.
\begin{equation}
\delta(\mbox{\boldmath{$X$}}) = \frac{1}{N(N-1)} \sum_{n \neq n'} \frac{1}{M} \sum_{m=1}^{M} |X_{nm}-X_{n'm}|.
\label{Ddef}
\end{equation}

The index $\varepsilon$ is the average difference between $X_{nm}$ and $1/M$. $X_{nm}=1/M$ for any $m$ means uniform diversification portfolio whose risk tends to be the smallest. The index $\varepsilon$ is defined by eq.(\ref{Ddef}). The relation $0 \leq \varepsilon \leq 1$ is satisfied. A more risk-exposed investment portfolio has a larger value of $\varepsilon$.
\begin{equation}
\varepsilon(\mbox{\boldmath{$X$}}) = \frac{1}{N} \sum_{m=1}^{M} |\sum_{n=1}^{N} X_{nm}-\frac{1}{M}|.
\label{defG}
\end{equation}

In case of a uniform diversification portfolio where $X_{nm}=1/M$, $\delta=0$ and $\varepsilon=0$. A bank-unique specialization portfolio means that individual banks invest in their own unique assets, $X_{nm_{{\rm U}}(n)}=1$. Generally, $m_{{\rm U}}(n) \neq m_{{\rm U}}(n')$ if $n \neq n'$. In this case, $\delta=1$ and $\varepsilon=0$. A system-wide specialization portfolio means that any banks invest in the same $m_{{\rm S}}$-th asset, $X_{nm_{{\rm S}}}=1$. In this case, $\delta=0$ and $\varepsilon=1$.

\subsection{Financial crisis scenario}

The trigger of the financial crisis in the ANWSER model is the fluctuation of the market prices of the assets. The fall in the market price of the unit quantity of the $m$-th asset is denoted by $v_{m}$. The loss which the $n$-th bank incurs is $e_{n} \sum_{m=1}^{M} X_{nm} v_{m}$. If this loss can not be absorbed by the equity capital, the bank goes bankrupt. The set of banks which go bankrupt in this initial stage is denoted by $\mbox{\boldmath{$F$}}_{0}$. It is given by eq.(\ref{defcond1}).
\begin{equation}
\mbox{\boldmath{$F$}}_{0} = \{n | c_{n} < e_{n} \sum_{m=1}^{M} X_{nm} v_{m} \}.
\label{defcond1}
\end{equation}

In the next stage, the insolvency of banks in $\mbox{\boldmath{$F$}}_{0}$ causes additional loss to the bank which lends any banks in $\mbox{\boldmath{$F$}}_{0}$. Any portions of the interbank loan are not paid off. If the total loss can not be absorbed by the equity capital, the bank goes bankrupt too. This is the transmission of financial distress and the basic mechanism of financial contagion. The set of banks which go bankrupt in this stage is denoted by $\mbox{\boldmath{$F$}}_{1}$. Similarly, this process goes on working many times until the financial contagion comes to a halt in the final stage. The set of banks which go bankrupt in the $j$-th stage $\mbox{\boldmath{$F$}}_{j}$ is given by eq.(\ref{defcond2}).
\begin{equation}
\mbox{\boldmath{$F$}}_{j} = \{n | c_{n} < e_{n} \sum_{m=1}^{M} X_{nm} v_{m} + \sum_{n' \in \mbox{\boldmath{$F$}}_{j-1}} l_{nn'} \}.
\label{defcond2}
\end{equation}

Formally, the set of banks which ended in bankruptcy in any stages is given by $\mbox{\boldmath{$F$}}_{\infty} = \bigcup_{j} \mbox{\boldmath{$F$}}_{j}$.

\subsection{Bankruptcy reproductive ratio}

The ANWSER model computes the bankruptcy reproductive ratio $A$. This quantifies the strength of a susequent chain effect as a measure of the risk of financial contagion. $A$ is defined by eq.(\ref{Adef}), which is the ratio of the final number of bankruptcies after the financial contagion to the initial number of bankruptcies immediately after the market prices fluctuate. It is a function of the macroscopic indices $\kappa$ and $\rho_{5}$ for a interbank credit network in \ref{interbank credit network}, and $\delta$ and $\varepsilon$ for an investment portfolio of banks in \ref{Investment portfolio}.
\begin{equation}
A(\kappa, \rho_{5}, \delta, \varepsilon) = \frac{|\mbox{\boldmath{$F$}}_{\infty}|}{|\mbox{\boldmath{$F$}}_{0}|}
\label{Adef}
\end{equation}

When $A=1$, financial contagion does not appear. As the value of $A$ increases, the financial contagion becomes more damaging. $A(\delta, \varepsilon)$ for given $\kappa$ and $\rho_{5}$ is called a risk landscape.

\section{Analytic solution for $N=2$}
\label{Analytic solution for $N=2$}

\subsection{Exact formula}

The ANWSER model in \ref{ANWSER model} can be solved analytically for a small financial system. The exact formula for the bankruptcy reproductive ratio is derived for $N=2$ and $M=2$. Both banks have the same amount of asset $a$. Both are creditors and debtors. The balance sheet becomes $l_{1}=b_{2}=\theta a$, $b_{1}=l_{2}=\theta a$, $e_{1}=e_{2}=(1-\theta)a$, $c_{1}=c_{2}=\gamma a$, and $d_{1}=d_{2}=(1-\gamma-\theta)a$. The fall in the market prices is a probability variable. It is assumed that $v_{1}$ and $v_{2}$ are independent and obey an exponential distribution in eq.(\ref{exponential}).
\begin{equation}
P(v_{m}) = 
\begin{cases}
\frac{\lambda}{2} e^{\lambda v_{m}} & {\rm if} \ v_{m} \leq 0\\
\frac{\lambda}{2} e^{-\lambda v_{m}} & {\rm if} \ v_{m} > 0
\end{cases}.
\label{exponential}
\end{equation}

\begin{figure}[!t]
\centering
\includegraphics[width=2.5in, angle=-90]{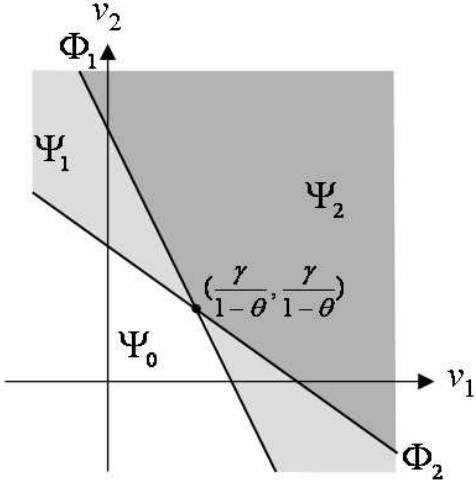}
\caption{Criterion for banks going bankrupt for $N=2$ and $M=2$ on the asset price plane ($v_{1}$, $v_{2}$). The boundaries $\Phi_{1}$ and $\Phi_{2}$ determine whether the two banks go bankrupt. None go bankrupt in the domain $\Psi_{0}$. Either of the two goes in $\Psi_{1}$. Both go in $\Psi_{2}$.}
\label{1124b2}
\end{figure}

Eq.(\ref{defcond1}) can be converted to the criterion for banks going bankrupt on the asset price plane ($v_{1}$, $v_{2}$). Fig.\ref{1124b2} shows the criterion. No banks go bankrupt in the domain $\Psi_{0}$. Either of the two banks goes bankrupt in $\Psi_{1}$. Both banks go bankrupt in $\Psi_{2}$. The boundaries $\Phi_{1}$ and $\Phi_{2}$ determine whether individual banks go bankrupt or not. They are given by eq.(\ref{Phi12}).
\begin{equation}
\begin{cases}
\Phi_{1}:\ \gamma = (1-\theta)\{X_{11} v_{1} + (1-X_{11}) v_{2}\}\\
\Phi_{2}:\ \gamma = (1-\theta)\{X_{21} v_{1} + (1-X_{21}) v_{2}\}
\end{cases}.
\label{Phi12}
\end{equation}

The probability of two banks going bankrupt is given by eq.(\ref{PF2}) where $\Lambda=\gamma \lambda/(1-\theta)$.
\begin{eqnarray}
&&p(|\mbox{\boldmath{$F$}}_{0}|=2) = \iint_{\Psi_{2}} P(v_{1}) P(v_{2}) {\rm d} v_{1} {\rm d} v_{2} \nonumber \\
&&\ \ \ = -\frac{(X_{11}-1)^{2}}{2(2X_{11}-1)} e^{\frac{\Lambda}{X_{11}-1}} + \frac{X_{21}^{2}}{2(2X_{21}-1)} e^{-\frac{\Lambda}{X_{21}}} \nonumber \\
&&\ \ \ -\frac{1}{4}(\frac{X_{11}-1}{2X_{11}-1} - \frac{X_{21}-1}{2X_{21}-1}) e^{-2\Lambda}.
\label{PF2}
\end{eqnarray}

The probability of one bank going bankrupt is given by eq.(\ref{PF1}).
\begin{eqnarray}
&& p(|\mbox{\boldmath{$F$}}_{0}|=1) = \iint_{\Psi_{1}} P(v_{1}) P(v_{2}) {\rm d} v_{1} {\rm d} v_{2} \nonumber \\
&&\ \ \ = \frac{(X_{11}-1)^{2}}{2(2X_{11}-1)} e^{\frac{\Lambda}{X_{11}-1}} + \frac{X_{11}^{2}}{2(2X_{11}-1)} e^{-\frac{\Lambda}{X_{11}}} \nonumber \\
&&\ \ \ - \frac{(X_{21}-1)^{2}}{2(2X_{21}-1)} e^{\frac{\Lambda}{X_{21}-1}} - \frac{X_{21}^{2}}{2(2X_{21}-1)} e^{-\frac{\Lambda}{X_{21}}} \nonumber \\
&&\ \ \ + \frac{1}{2}(\frac{X_{11}-1}{2X_{11}-1} - \frac{X_{21}-1}{2X_{21}-1}) e^{-2\Lambda}.
\label{PF1}
\end{eqnarray}

The probability of no banks going bankrupt is given by eq.(\ref{PF0}).
\begin{eqnarray}
&& p(|\mbox{\boldmath{$F$}}_{0}|=0) = \iint_{\Psi_{0}} P(v_{1}) P(v_{2}) {\rm d} v_{1} {\rm d} v_{2} \nonumber \\
&&\ \ \ = - \frac{X_{11}^{2}}{2(2X_{11}-1)} e^{-\frac{\Lambda}{X_{11}}} + \frac{(X_{21}-1)^{2}}{2(2X_{21}-1)} e^{\frac{\Lambda}{X_{21}-1}} \nonumber \\
&&\ \ \ - \frac{1}{4}(\frac{X_{11}-1}{2X_{11}-1} - \frac{X_{21}-1}{2X_{21}-1}) e^{-2\Lambda} +1.
\label{PF0}
\end{eqnarray}

\begin{figure}[!t]
\centering
\includegraphics[width=2.5in, angle=-90]{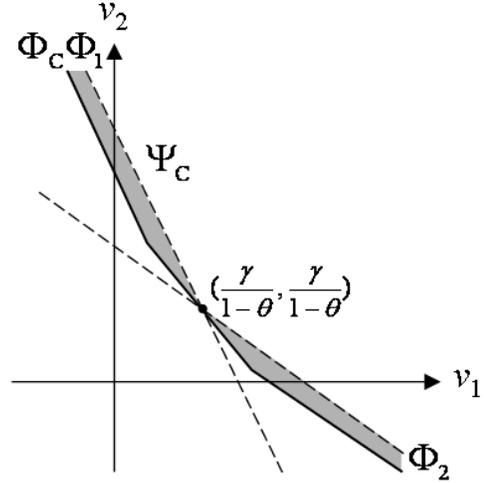}
\caption{Criterion for financial contagion for $N=2$ and $M=2$ on the asset price plane ($v_{1}$, $v_{2}$). If one bank went bankrupt, the other bank goes bankrupt too in the domain $\Psi_{{\rm C}}$. The boundaries of $\Psi_{{\rm C}}$ are $\Phi_{1}$, $\Phi_{2}$, and $\Phi_{{\rm C}}$.}
\label{1124b3}
\end{figure}

Eq.(\ref{defcond2}) can be converted to the criterion for financial contagion on the asset price plane. Fig.\ref{1124b3} shows the criterion. If one bank goes bankrupt, the other bank goes bankrupt too in the domain $\Psi_{{\rm C}}$. The boundaries of $\Psi_{{\rm C}}$ are $\Phi_{1}$, $\Phi_{2}$, and $\Phi_{{\rm C}}$. $\Phi_{{\rm C}}$ is given by eq.(\ref{PhiC}).
\begin{equation}
\Phi_{{\rm C}}:
\begin{cases}
\gamma = \min( (1-\theta)\{X_{21}v_{1}+(1-X_{21})v_{2}\} - \gamma, \theta) \\
\ \ +(1-\theta)\{X_{11}v_{1}+(1-X_{11})v_{2}\} \ {\rm if} \ v_{1} \leq \frac{\gamma}{1-\theta} \\
\gamma = \min( (1-\theta)\{X_{11}v_{1}+(1-X_{11})v_{2}\} - \gamma, \theta) \\
\ \ +(1-\theta)\{X_{21}v_{1}+(1-X_{21})v_{2}\} \ {\rm if} \ v_{1} > \frac{\gamma}{1-\theta}
\end{cases}.
\label{PhiC}
\end{equation}

The probability of financial contagion is given by eq.(\ref{PFC}). The exact formula for eq.(\ref{PFC}) can be derived although it is complicated.
\begin{eqnarray}
p(|\mbox{\boldmath{$F$}}_{0}|=1, |\mbox{\boldmath{$F$}}_{1}|=2) = \iint_{\Psi_{{\rm C}}} P(v_{1}) P(v_{2}) {\rm d} v_{1} {\rm d} v_{2}.
\label{PFC}
\end{eqnarray}

The exact formula for the bankruptcy reproductive ratio is derived from the probabilities in eq.(\ref{PF2}) through (\ref{PFC}) although it is also complicated. For $M=2$, $\delta=|X_{11}-X_{21}|$, $\varepsilon=|X_{11}+X_{21}-1|$, and $0 \leq \delta+\varepsilon \leq 1$ are satisfied. The bankruptcy reproductive ratio is obtained as a function of $\delta$ and $\varepsilon$, as well as a function of $X_{11}$ and $X_{21}$. This is the risk landscape $A(\delta, \varepsilon)$.

\subsection{Risk landscape}
\label{Risk lanscape N=2}

\begin{figure}[!t]
\centering
\includegraphics[width=2.5in, angle=-90]{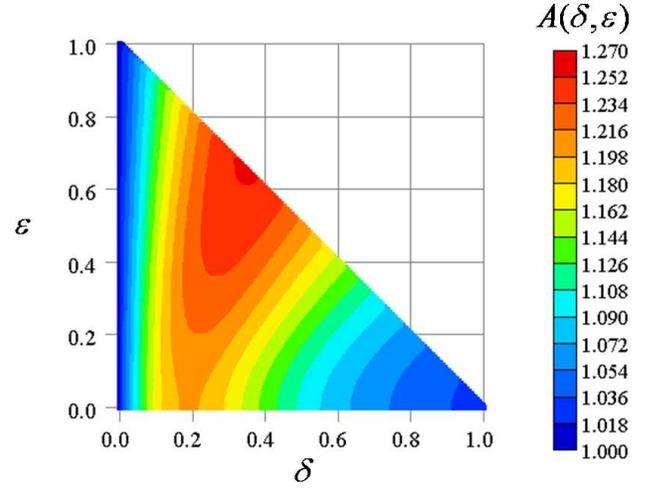}
\caption{Risk landscape derived analytically by the ANWSER model. The bankruptcy reproductive ratio $A$ is drawn as a function of the diversity $\delta$ and risk exposure $\varepsilon$ when $\theta=0.1$ and $\gamma=0.05$ for $N=2$ and $M=2$. The ratio is measured at the average point of the number of bankruptcies.}
\label{1124b5}
\end{figure}

Fig.\ref{1124b5} shows the risk landscape $A(\delta, \varepsilon)$ which is measured at the average point of the number of bankruptcies. The value of $\lambda$ in eq.(\ref{exponential}) is adjusted so that the probability of bankruptcy of an individual bank having $\gamma=0.07$ can be $10^{-3}$. The bankruptcy reproductive ratio becomes large when the value of $\delta$ is 0.2 through 0.4. Although the condition under which the risk of financial contagion vanishes ($A=1$) is not obtained analytically, Fig.\ref{1124b5} demonstrates that large diversity among banks in the investment portfolio reduces the risk of financial contagion. A bank-unique specialization portfolio is more suitable than a uniform diversification portfolio and a system-wide specialization in strengthening the robustness of a finacial system.

\begin{figure}[!t]
\centering
\includegraphics[width=2.6in, angle=-90]{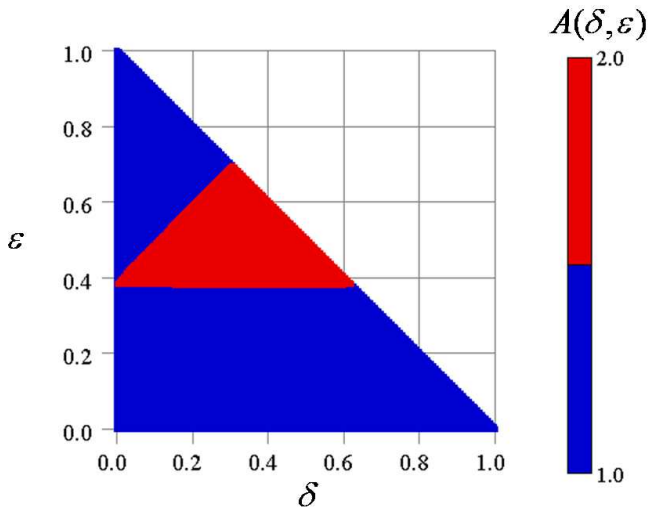}
\caption{Risk landscape derived analytically by the ANWSER model. The bankruptcy reproductive ratio $A$ is drawn as a function of the diversity $\delta$ and risk exposure $\varepsilon$ when $\theta=0.1$ and $\gamma=0.05$ for $N=2$ and $M=2$. The ratio is measured at the 999th 1000-quantile point (the worst case) of the number of bankruptcies.}
\label{1124b6}
\end{figure}

Fig.\ref{1124b6} shows $A(\delta, \varepsilon)$ which is measured at the 999th 1000-quantile point. The bankruptcy reproductive ratio is either 1 or 2. Financial contagion disappears when $\delta > 0.6$. Again, large diversity reduces the risk of financial contagion. Small risk exposure is also relevant in eliminating the financial contagion.

\section{Numerical solution for $N=500$}
\label{Numerical solution for $N=500$}

\subsection{Simulation experiment}

\begin{figure}[!t]
\centering
\includegraphics[width=2.3in, angle=-90]{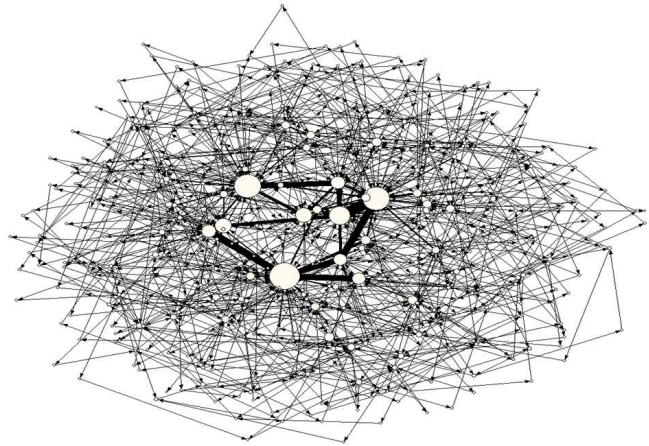}
\caption{Heterogeneous interbank credit network for $N=500$. The size of a vertex represents the amount of asset of a bank. The width of an edge represents the amount of an interbank loan between the banks at its ends.}
\label{1124b4}
\end{figure}

The ANWSER model in \ref{ANWSER model} can be solved numerically with a Monte-Carlo simulation technique for a large financial system. Given $\kappa$ and $\rho_{5}$, a number of samples for a heterogeneous interbank credit network $\mbox{\boldmath{$T$}}$ are generated randomly. Fig.\ref{1124b4} shows an example of a heterogeneous interbank credit network when $N=500$. The network is a generalized Barab\'{a}si-Albert model\cite{Dorogovtsev}, \cite{Barabasi}. This is a random graph with the mechanism of growth and preferential attachment, which becomes scale-free as $N$ goes to infinity, that is, the distribution of the degree $k$ obeys the power law, $P(k) \propto k^{-\alpha}$ where $\alpha \geq 2$. There is a significant probability of the presence of very big banks. This is the origin of heterogeneity.

Given $\delta$ and $\varepsilon$, a number of samples for an investment portfolio $\mbox{\boldmath{$X$}}$ are generated randomly. It is assumed that the fall in the market prices $v_{m}$ are independent and obey a Student t-distribution whose degree of freedom is 1.5. This is a long-tailed distribution. It is suitable to describe a sudden large fluctuation of market prices. A number of samples for $v_{m}$ are generated randomly. The number of bankruptcies at the average point and the 999th 1000-quantile point can be obtained from the frequency distribution of those samples.

\subsection{Risk landscape}

\begin{figure}[!t]
\centering
\includegraphics[width=2.6in, angle=-90]{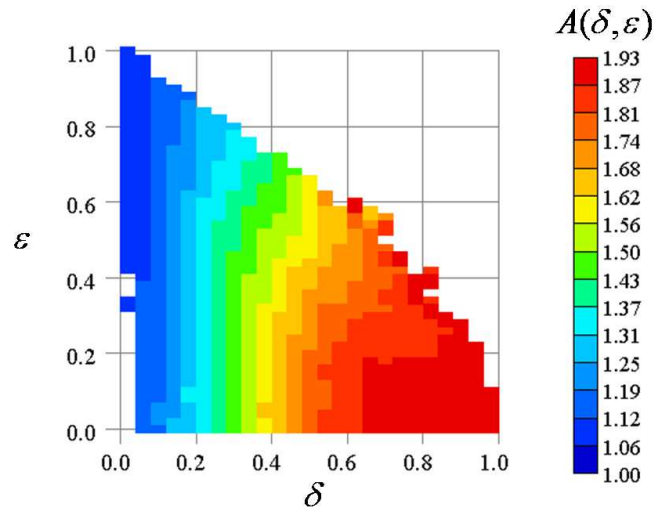}
\caption{Risk landscape obtained numerically by the ANWSER model. The bankruptcy reproductive ratio $A$ is drawn as a function of the diversity $\delta$ and risk exposure $\varepsilon$ when $\theta=0.1$, $\gamma=0.07$, $K=25$, and $\rho_{5}=0.25$ for $N=500$ and $M=2$. The ratio is measured at the average point of the number of bankruptcies.}
\label{1124b7}
\end{figure}

Fig.\ref{1124b7} shows the risk landscape $A(\delta, \varepsilon)$ which is measured at the average point of the number of bankruptcies. The conditions $N=500$, $\kappa=25$, and $\rho_{5}=0.25$ are equivalent to those for BOJ-Net. The bankruptcy reproductive ratio becomes larger as $\delta$ approaches 1. The largest value of $A$ is 1.9. This landscape is quite different from that in Fig.\ref{1124b5}. Risk exposure does not have a strong impact on the bankruptcy reproductive ratio.

\begin{figure}[!t]
\centering
\includegraphics[width=2.6in, angle=-90]{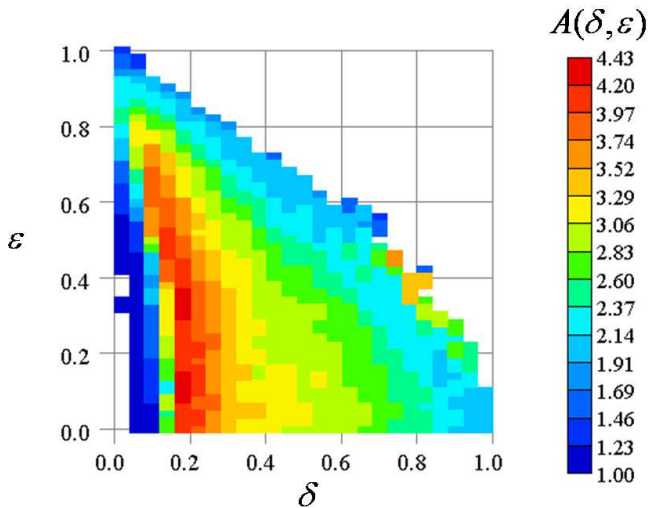}
\caption{Risk landscape obtained numerically by the ANWSER model. The bankruptcy reproductive ratio $A$ is drawn as a function of the diversity $\delta$ and risk exposure $\varepsilon$ when $\theta=0.1$, $\gamma=0.07$, $K=25$, and $\rho_{5}=0.25$ for $N=500$ and $M=2$. The ratio is measured at the 999th 1000-quantile point (the worst case) of the number of bankruptcies.}
\label{1124b8}
\end{figure}

Fig.\ref{1124b8} shows $A(\delta, \varepsilon)$ which is measured at the 999th 1000-quantile point. The bankruptcy reproductive ratio becomes quite large when the value of $\delta$ is 0.2. The largest value of $A$ is 4.4. Large diversity among banks in the investment portfolio reduces the risk of financial contagion essentially. The results are striking. Large diversity among banks makes financial contagion more damaging on the average. A uniform diversification portfolio and a system-wide specialization portfolio are more suitable than a bank-unique specialization portfolio. But large diversity is essentially effective in eliminating the risk of financial contagion in the worst case of financial crisis scenarios. A bank-unique specialization portfolio is more suitable than a uniform diversification portfolio and a system-wide specialization portfolio in strengthening the robustness of a financial system.

\section{Conclusion}

One of the lessons of the financial crisis in 2007 is that supervisors and other relevant authorities failed to capture the systemic risk hidden behind banks just by keeping a close eye on individual banks and economic fundamentals. The scope of the recent arguments on the capital surcharge is still the resilience of individual global systemically important banks. The analysis with the ANWSER model, however, demonstrates that the risk of financial contagion depends on the macroscopic indices, diversity and risk exposure of an investment portfolio of banks. Large diversity is essentially effective in eliminating the risk of financial contagion in the worst case of financial crisis scenarios. A bank-unique specialization portfolio is more suitable than a uniform diversification portfolio and a system-wide specialization portfolio. These results form the basis for an effective design principle for supervisors and other relevant authorities in making bank regulatory policies to strengthen the robustness of a financial system.

The ANWSER model will be extended to address a more practical finance system. The model will include settlement dates (overnight, short-term, or long-term) of interbank loans, withdrawal and netting of interbank loans, refinancing and other means to raise money, liquidity of assets, correlation between the market prices of assets, and other market mechanisms. The bank run in a fractional reserve banking when a financial crisis precipitates a banking panic is also of interest as a topic for future works. The market price of an external asset is a linear combination of independent and identically distributed probability variables. If the number of assets increases, the price is less volatile, the loss which a bank incurs in the initial stage is less fluctuating, and the number of bankruptcies tends to decrease. But it is not evident whether the number of assets has a big impact on the bankruptcy reproductive ratio and the risk landscape. This topic is also for future works.

Another issue is that the interbank credit network and the investment porfolio are not static in reality and can not be observed directly either. In the study of epidemic contagion in social network analysis, the transportation network and relevant parameters are inferred from the observation on the number of patients with a statistical analysis\cite{Maeno11}, \cite{Maeno10}. Then, an impending epidemic outbreak may be predicted with a Monte-Carlo simulation technique. Similarly, the prediction of the risk of a financial system may be founded on the facts which are obtained by observing individual banks and inferring the interbank credit network and the investment porfolio. Such prediction aids the supervisors and other relevant authorities in designing the financial system theoretically and in making regulatory policies in practice. This is the goal of an emerging field of systems economics.

\section*{Acknowledgment}
The authors would like to thank Hidetoshi Tanimura, Ernst\&Young ShinNihon LLC, and Kenichi Amagai, Institute for International Socio-economic Studies, for their advice and discussion.

\end{document}